\documentclass[reprint,prl]{revtex4-1}
\usepackage[english]{babel}
\usepackage{amstext}
\usepackage{amssymb}
\usepackage{graphicx}
\usepackage{esint}

\usepackage{babel}
\begin{document}

\title{A rare event algorithm links transitions in turbulent flows with activated nucleations}

\author{Freddy Bouchet}\email{Freddy.Bouchet@ens-lyon.fr}\author{Joran Rolland}\email{Joran.Rolland@ens-lyon.fr}\affiliation{Univ Lyon, Ens de Lyon, Univ Claude Bernard Lyon 1, CNRS, Laboratoire de Physique, F-69342 Lyon, France. }
\author{Eric Simonnet}\email{eric.simonnet@inphyni.cnrs.fr}\affiliation{InPhyNi, UMR CNRS 7010, Nice, France.}



\begin{abstract}
Many turbulent flows undergo drastic and abrupt configuration changes with huge impacts. As a paradigmatic example we study the multistability of jet dynamics in a barotropic beta plane model of atmosphere dynamics. It is considered as the Ising model for Jupiter troposphere dynamics. Using the adaptive multilevel splitting, a rare event algorithm, we are able to get a very large statistics of transition paths, the extremely rare transitions from one state of the system to another. This new approach opens the way for addressing a set of questions that are out of reach through direct numerical simulations. We demonstrate for the first time the concentration of transition paths close to instantons, in a numerical simulation of genuine turbulent flows. We show that the transition is a noise-activated nucleation of vorticity bands. We address for the first time the existence of Arrhenius laws in turbulent flows. The methodology we developed shall prove useful to study many other transitions related to drastic changes for the turbulent dynamics of climate, geophysical, astrophysical and engineering applications. This opens a new range of studies impossible so far, and bring turbulent phenomena in the realm of non-equilibrium statistical mechanics. 
\end{abstract}

\maketitle

Many turbulent flows undergo drastic and abrupt configuration
changes with huge impacts \cite{Sommeria1986,Ravelet_Marie_Chiffaudel_Daviaud_PRL2004,Berhanu2007,wei2015multiple,zimmerman2011bi,huisman2014multiple,pomeau2016long}.
The Earth magnetic field reverses on geological time scales due to
turbulent motion of the Earth's metal core \cite{Berhanu2007},
wall flows transition from laminar to turbulent \cite{pomeau2016long,barkley2015rise,shih2015ecological},
experiments in convection turbulence show bistability \cite{wei2015multiple,zimmerman2011bi,huisman2014multiple},
global climate changes like the glacial-interglacial transitions or
the Dansgaard\textendash Oeschger events are related to the turbulent
oceans and atmosphere coupled to ice and carbon dioxide dynamics \cite{Rahmstorf_2002_Nature}. Is  the kinetics and phenomenology of these turbulent transitions analogous
to phase transitions in condensed matter, and rare conformational
changes of molecules in chemistry and biochemistry? These key questions
have not been addressed so far because of the difficulty related to
the numerical complexity: We need both a proper turbulence representation
and run extremely long simulations to observe transitions. In this letter we
show that a new numerical approach based on rare event algorithms
improves exponentially our capabilities. With this tool, we make the first numerical study of metastability and spontaneous transitions for a genuine turbulent dynamics. We study atmospheric turbulent
jet transitions, relevant to describe abrupt climate changes on Jupiter's
troposphere.

Jupiter pictures 
show fascinating zonal bands whose color
are correlated with the troposphere flow vorticity.
 Those bands correspond
to East-West (zonal) velocity jets, which are stationary for centuries.
During the period 1939-1940 a fantastic phenomenon occurred: the planet
lost one of its jets \cite{rogers1995giant}, which was replaced by
three white anticyclones. Phil Marcus subsequently called this event
a Jupiter's sudden climate change \cite{marcus2004prediction}. This
rare event is one example among thousands of sudden transitions betwee
attractors in the self-organization of billions of vortices in turbulent
flows. In this letter, we study the barotropic beta-plane quasi-geostrophic
equations: the simplest model that describes the turbulent atmosphere
jet self-organization \cite{VAL,Dritschel_McIntyre_2008JAtS,galperin2001universal,Farrell_Ioannou_JAS_2007,tobias2013direct,Srinivasan-Young-2011-JAS}.
It is the Ising model of atmosphere dynamics: the simplest model to
describe jet formation, although too simple to be quantitatively realistic.

\begin{figure}
\centerline{\includegraphics[width=0.4\textwidth]{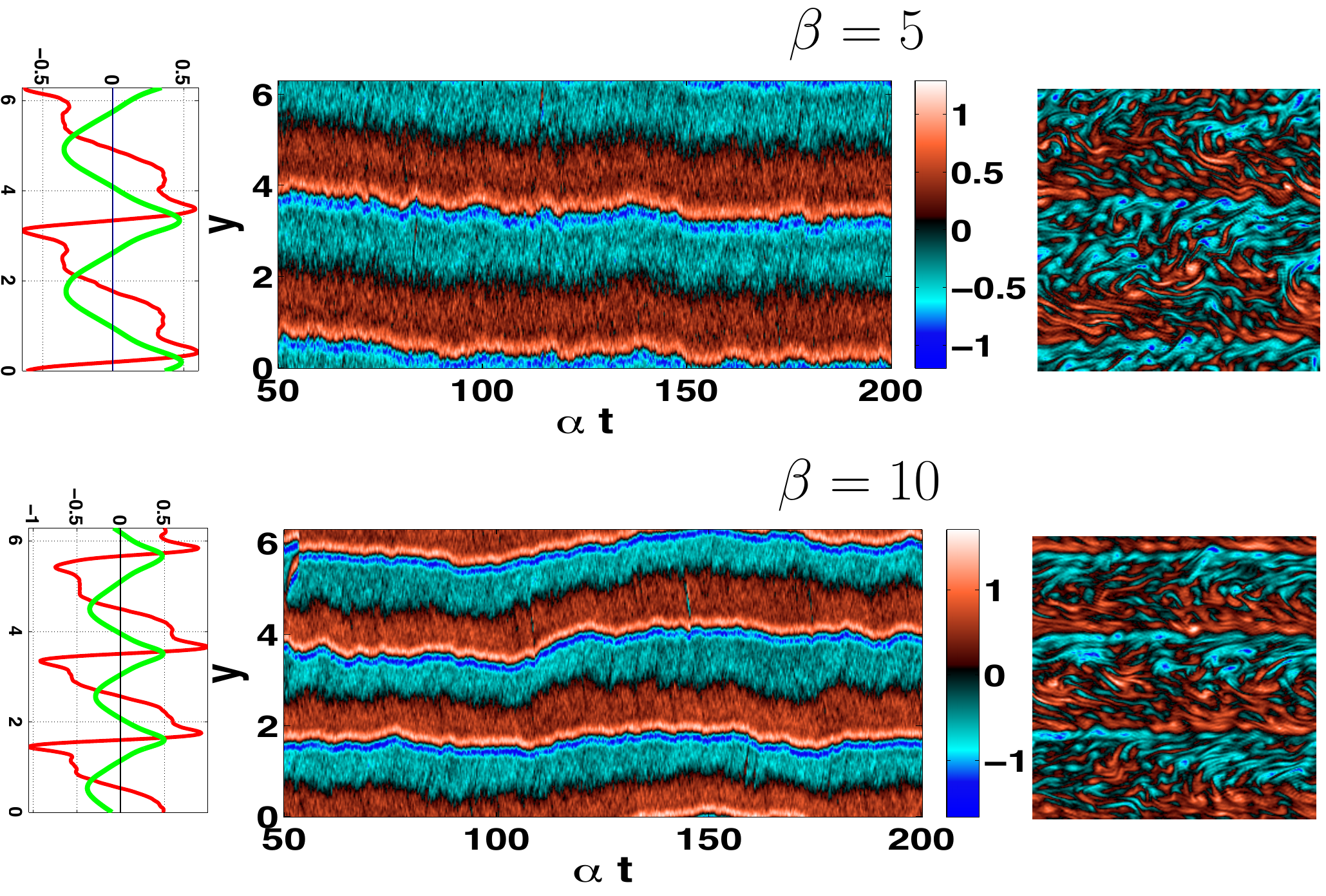}}
\caption{ Right panels:
typical snapshots of the vorticity fields (the colors show the vorticity
values, with red for positive ones, blue for negative ones and black
for intermediate ones; the range of vorticity is {[}-1,1{]}). Middle
panels: Hovm\"oller diagrams of zonally averaged vorticity (the horizontal
axis is $\alpha t$, the vertical one is $y$, and the color represents
the $x-$averaged vorticity, $\alpha=1.20\ 10^{-3}$). Left panels:
time and zonally averaged vorticity (red) and velocity (green). The
top plots show a two jet state for $\beta=5$, while the bottom ones
show a three jet one for $\beta=10$.}
\label{fig1}
\end{figure}

The dimensionless version of the model equations read
\begin{equation}
\partial_{t}\omega+\mathbf{v}\cdot\mathbf{\nabla}\omega+\beta v_{y}=-\alpha\omega-\nu_{n}\left(-\Delta\right)^{n}\omega+\sqrt{2\alpha}\eta,\label{eq:barotropic}
\end{equation}
where $\mathbf{v}=\mathbf{e}_{z}\times\mathbf{\nabla}\psi$ is the
non-divergent velocity, $v_{y}$ the North-South velocity component,
$\omega=\Delta\psi$ and $\psi$ are the vorticity and the streamfunction,
respectively, $\alpha$ is a linear friction coefficient, see Supplemental Material for the dimensional equations. The noise
$\eta$ forces the flow dynamics and is precisely defined in the Supplemental Material file.
When $\beta=0$, those equations are the two-dimensional stochastic Navier--Stokes equations for which a few rare transitions have been observed in the past between dipole and jet states \cite{Bouchet_Simonnet_2008}, and for which impressive explicit relation between the energy injection rate and the Reynolds stresses have been recently derived \cite{laurie2014universal,kolokolov2016structure,kolokolov2016velocity}. Such relations have been further justified and extended to the case $\beta \neq 0$ \cite{woillez2017theoretical}, see also \cite{Srinivasan-Young-2011-JAS}. For $\beta \neq 0$  and for small enough $\alpha$
the vorticity dynamics self-organizes into zonal bands like on Jupiter
(Fig.~\ref{fig1}). The dimensionless parameter $\beta$ is proportional
to $\beta_{d}$ that measures the local variations of the Coriolis
parameter, and is related to the Rhines scale, defined in the Supplementary Materials. The number
of jets roughly scales like $\beta^{1/2}$ \cite{Dritschel_McIntyre_2008JAtS}.
Fig.~\ref{fig1} shows metastable turbulent states with either two or three
alternating jets, for two different values of $\beta$.

\begin{figure}
\centerline{\includegraphics[width=0.5\textwidth]{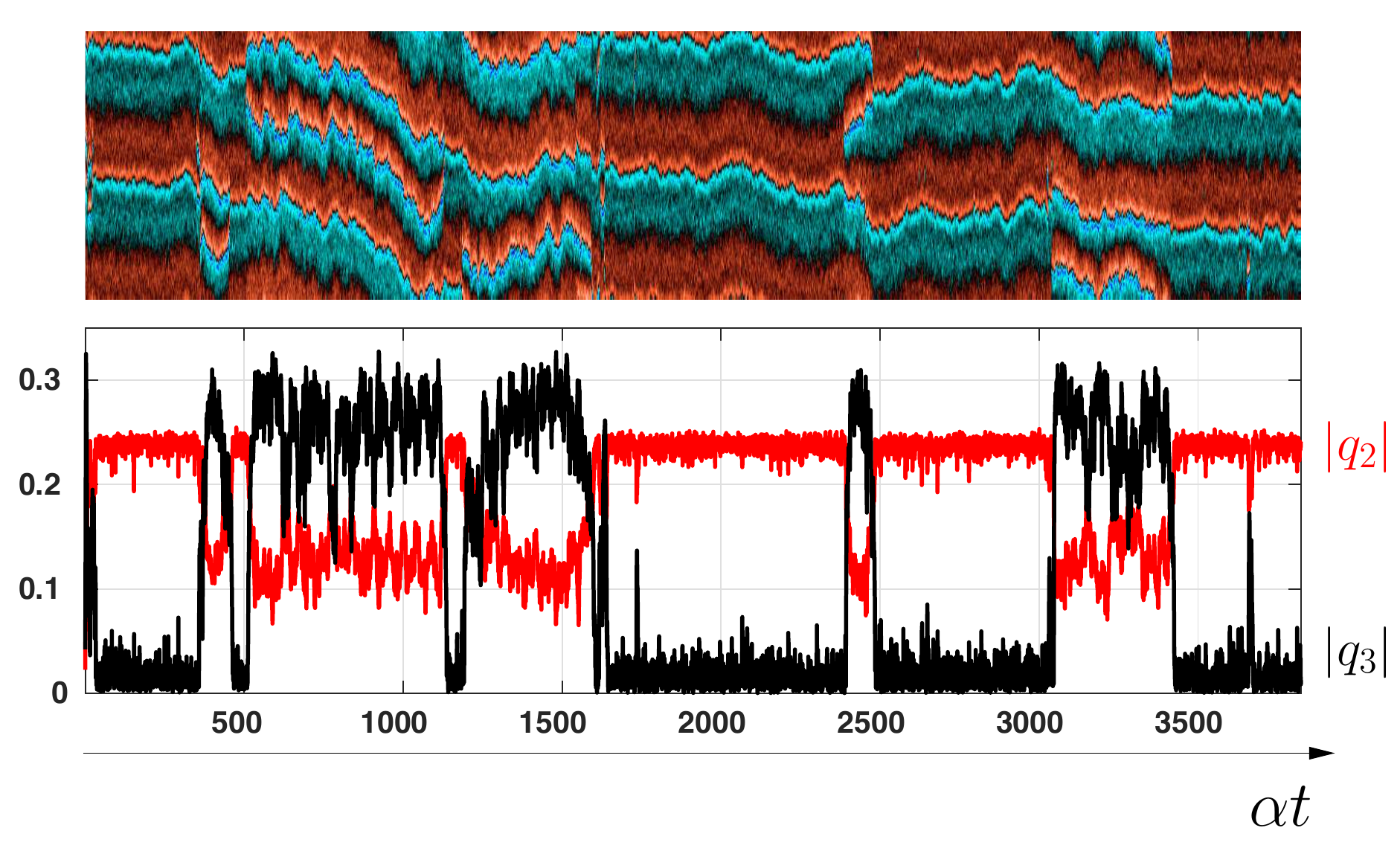}}
\caption{
 Rare transitions between
turbulent attractors with respectively 2 and 3 alternating jets. Upper
panel: Hovm\"oller diagram of the zonal mean vorticity. Lower panel:
timeseries of the modulus of $q_{2}$ (red) and $q_{3}$ (black),
the zonal mean vorticity Fourier components, for wavenumber 2 and
3 respectively, versus the rescaled time $\alpha t$ ($\alpha=1.2\cdot10^{-3}$
and $\beta=5.26$).}
\label{fig2}
\end{figure}

When $\beta$ is increased, one expects to see transitions from attractors
with $2$ to $3$ alternating jets ($2\rightarrow3$ transitions).
As there is no related symmetry breaking, one may expect these transitions
to be first order ones with discontinuous jumps of some order parameters.
In situations with discontinuous transitions when an external parameter
$\beta$ is changed, one expects for each bifurcation a multistability
range $(\beta_{1},\beta_{2})$ in which two (or more) possible states
exist for a single value of $\beta$. Such a bistability has indeed
been observed \cite{Farrell_Ioannou_JAS_2007}, by changing the model
initial conditions. Fig.~\ref{fig2} shows for the first time spontaneous transitions
between the two bistable states. The transitions are well characterized
by the Fourier components $q_{n}=\int\mbox{d}x\mbox{d}y\,\omega(x,y)\mbox{e}^{iny}/(2\pi)^{2}$
for $n=2$ and $n=3$: Fig.~\ref{fig2} features five $2\to3$ transitions,
and five $3\to2$ transitions in about $10^{6}$ turnover times.

\begin{figure}
\centerline{\includegraphics[width=0.45\textwidth]{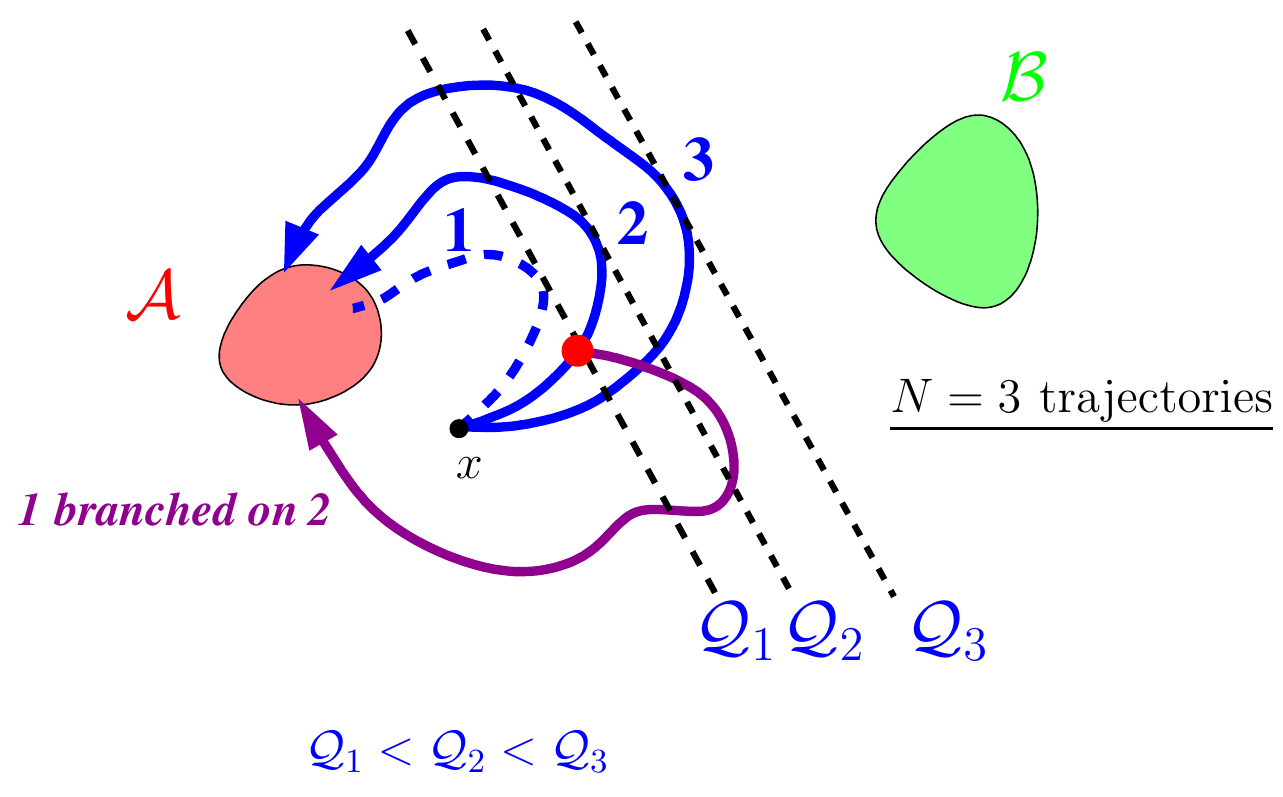}}
\caption{(a) Sketch of \textbf{the Adaptive Multilevel Splitting (AMS) algorithm},
one of the most versatile rare event algorithms. The aim of the algorithm
is to compute the very small probability to go from a set $\mathcal{A}$ (for instance
a two jet state) to a set $\mathcal{B}$ (for instance a three jet state). $N$
initial trajectories are computed from the model ($N$ is typically
a few thousands). A score function $\mathcal{Q}$ measures how far each trajectory
goes in the direction of $\mathcal{B}$. The worst trajectory is deleted (trajectory
No. 1 on the sketch). It is replaced by a new trajectory (purple trajectory)
whose initial condition is picked from one of the other previous trajectories
(trajectory 2 on the sketch) at the time (red dot on the sketch) when it was crossing the $\mathcal{Q}$ level with the maximum value of $\mathcal{Q}$ for the deleted trajectory. This last step is called resampling or
cloning. As the new set of $N$ trajectories has been obtained by selecting $N-1$
trajectories among $N$, and computing a new trajectory which has the same probability as the $N-1$ other ones, the new set has a probability $1-1/N$. The resampling
step is iterated K times, leading to new trajectory set with probability
$\left(1-1/N\right)^{K}$. This very efficiently produces extremely
rare transitions from one attractor to another and gives an unbiased
estimate of their probability (see the Supplementary Material file for more precise explanations).}
\label{fig3}
\end{figure}

We would like to address the following basic questions: What are the
transition rates? What are the relative probability of each attractors
(equilibrium constants)? Do the transition trajectories concentrate
close to instantons like in statistical physics \cite{Bouchet_Nardini_Tangarife_Kazimierz}?
Do the thousands of small scale vortices act collectively as atoms
in a nucleation process? Unfortunately, such questions are unaffordable
using direct numerical simulations. Observing such rare spontaneous
transitions in turbulent flows is highly unusual as most turbulent
simulations last a few turnover times at most, because of the huge numerical
cost. This limitation is a wall that drastically limits the study
of transitions in turbulent flows to extremely simple models and a
few transitions only. In order to study rare transitions in turbulent
flows, we consider a completely new approach in turbulence studies:
using the adaptive multilevel splitting algorithm \cite{CG07,Brehierlelievrerousset,Rolland_Bouchet_Simonnet_JStatPhys_2016}
(see Fig.~\ref{fig3}). This rare event algorithm belongs to the
family of splitting algorithms, where an ensemble of trajectories
are simulated and subjected to a succession of selections, cloning
or killing, and dynamical mutation steps. The principle of the algorithm \cite{CG07} is
described in the legend of Fig.~\ref{fig3}. A full description of the algorithm
and of the method to compute transition rates is described in \cite{Rolland_Bouchet_Simonnet_JStatPhys_2016}. Its mathematical properties (convergence, fluctuations, etc) have been studied recently \cite{Brehierlelievrerousset,cerouguyader2016}.
This algorithm has first been tested in extremely simple dynamics
with few degrees of freedom \cite{CG07}. It has been applied for
the first time to  a partial differential equation, the Ginzburg--Landau dynamics, in \cite{Rolland_Bouchet_Simonnet_JStatPhys_2016}.
In \cite{Rolland_Bouchet_Simonnet_JStatPhys_2016}, for the equilibrium Ginzburg--Landau dynamics, a very precise
comparison of the AMS algorithm results with explicit analytic results
of the Freidlin-Wentzell theory is performed, showing that the algorithm
can faithfully compute averaged transition times of order of $10^{15}$ larger
than the typical duration of a direct numerical simulation. This letter
describes the first application of the adaptive multilevel splitting
algorithm to turbulent flows, and to complex non equilibrium dynamics, where analytical results are out of reach.
This is also the first use of a rare event algorithm to study transitions
in turbulence that can not be studied through direct numerical simulations. Using this algorithm we
have been able to compute thousands of spontaneous transitions and
their probability. Table~\ref{tab1} shows the exponential reduction
of computational time in order to compute thousands of transitions.

\begin{figure}
\centerline{\includegraphics[width=0.45\textwidth]{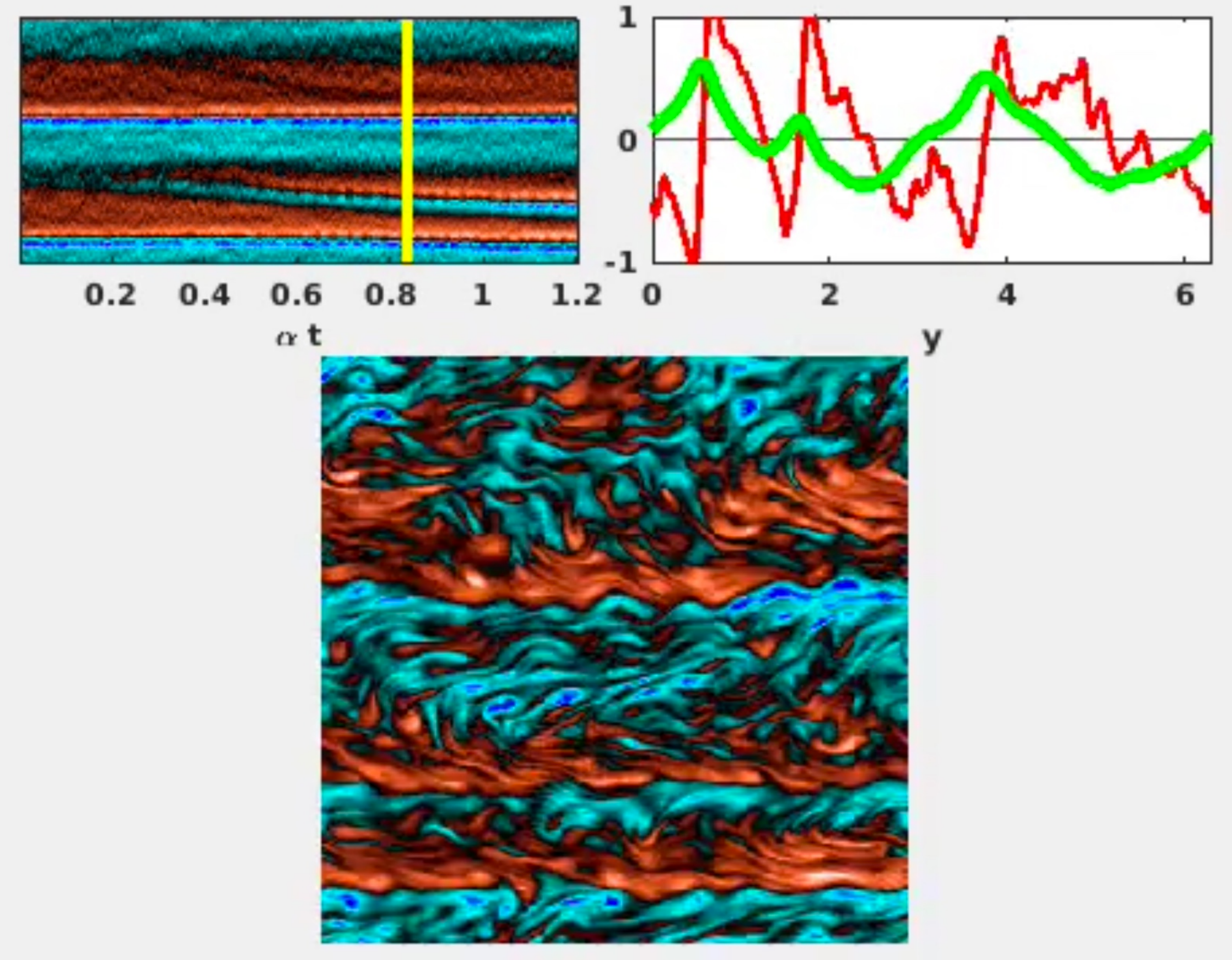}
}
\caption{\textbf{Nucleation of a new jet.}
The panels display the same quantities as Fig.~\ref{fig1}. A set of vortices
is able to nucleate a new band of blue vorticity, a very unlikely
process, leading to the birth of a new jet seen on the green velocity
curve. As in nucleation processes in condensed matter, once the nucleated
structure is large enough the new jet will be stable and persist for
extremely long times, as seen on the Hovm\"oller diagram (see also the Supplementary video).}
\label{fig4}
\end{figure}

\begin{figure}
\centerline{(a)\includegraphics[width=0.3\textwidth]{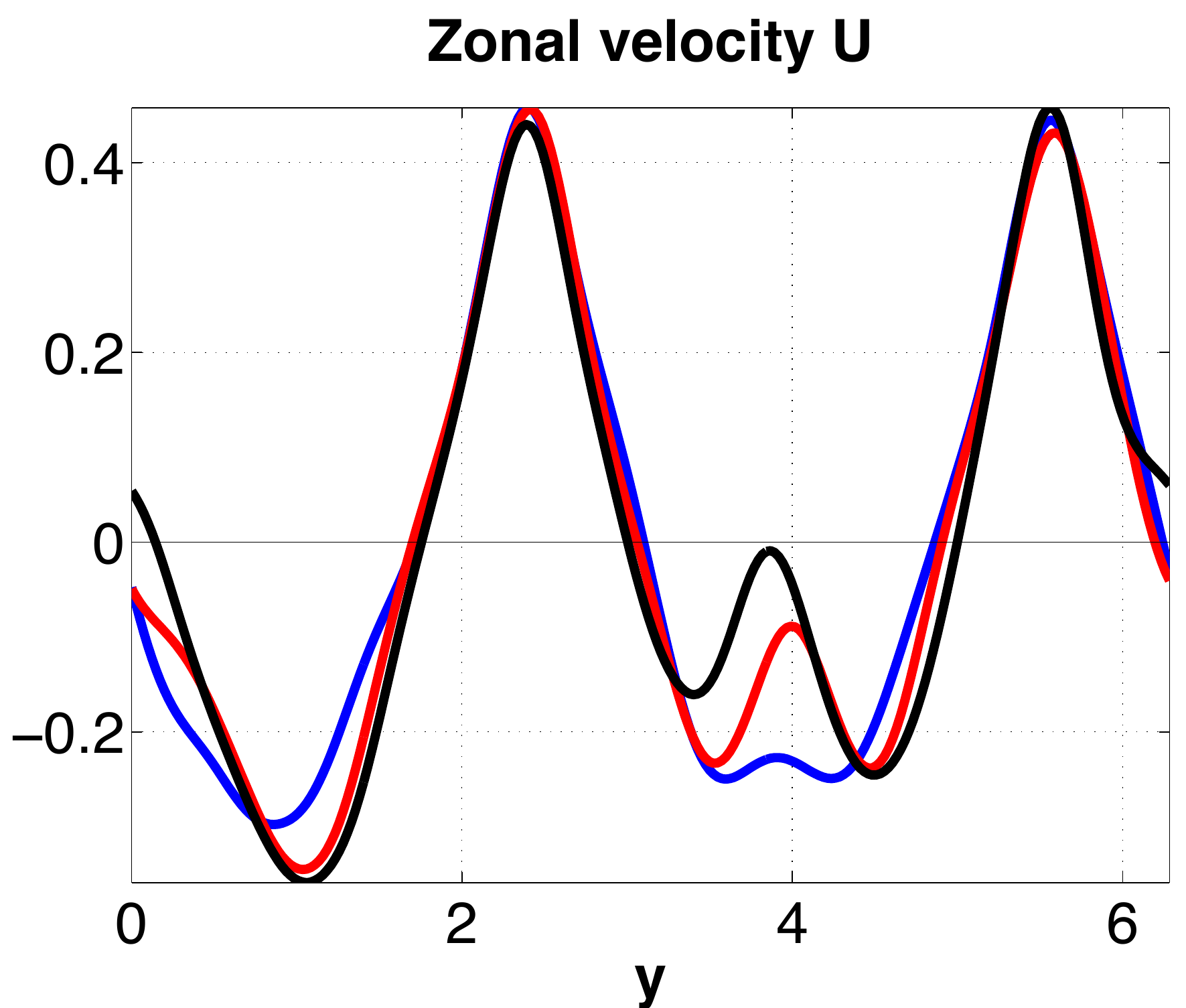}}
\centerline{(b)\includegraphics[width=0.3\textwidth]{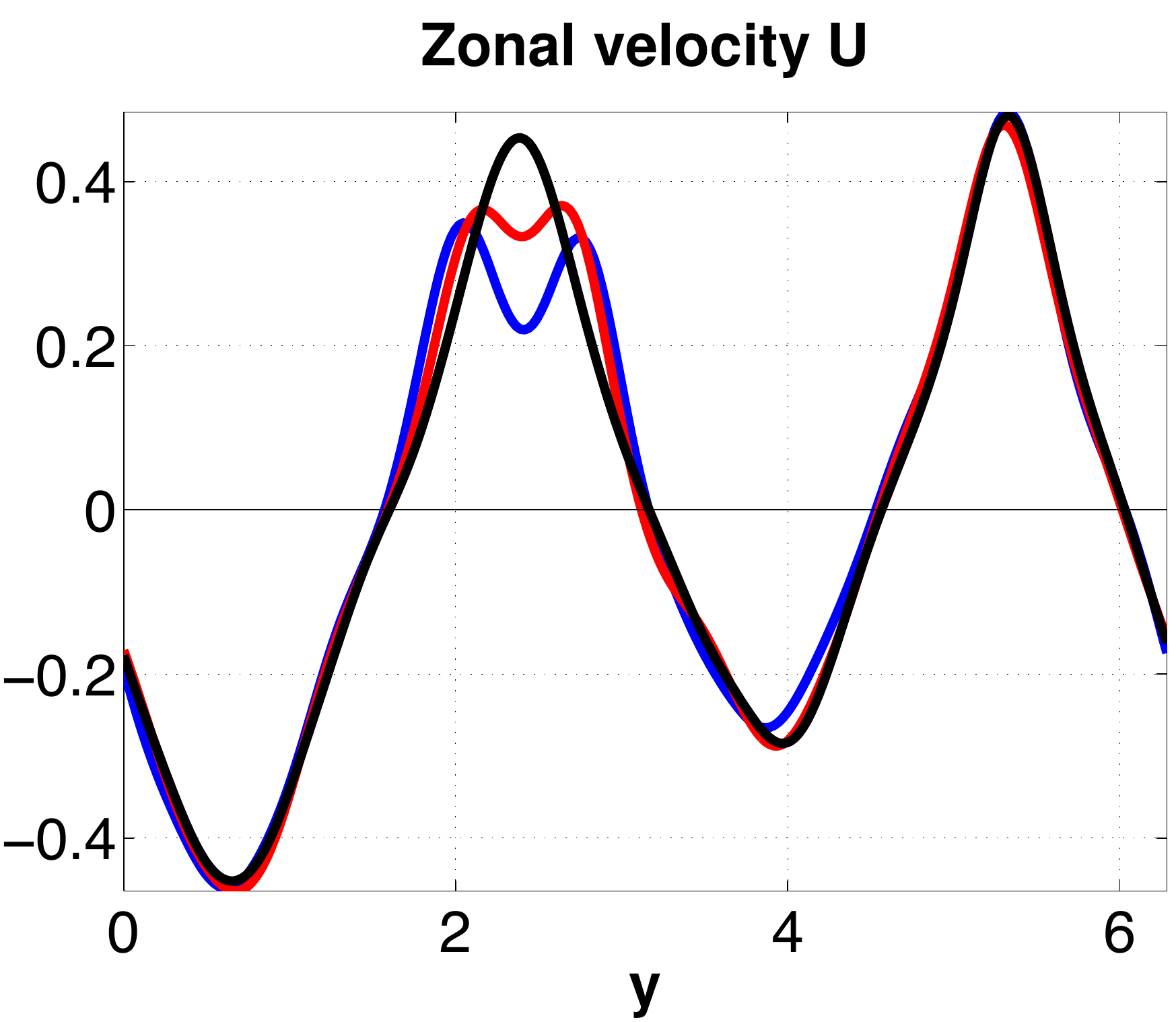}}
\caption{
(a) Zonally averaged velocity during
the nucleation of a third jet in a $2\to3$ jet transition: The blue,
red and black curves show the velocity field at the start, at an intermediate
stage, and at a more advanced stage of the nucleation, respectively.
(b) Zonally averaged velocity during the coalescence of two jets in
a $3\to2$ transition: The blue, red and black curves show the
velocity field before, just before, and after the merging, respectively.
These 4 plots illustrate that while transitions appear at random time,
their dynamic is predictable. }
\label{fig5}
\end{figure}

Fig.~\ref{fig4} and the associated movie (Supplementary Video) describe $2\to3$
transitions for $\alpha=6.0\ 10^{-4}$. Both the movie and the figure
clearly illustrate that a new jet formation proceeds through the nucleation
of two new ensembles of small positive and negative vortices lying
in an area of overall zero vorticity located at a westward jet. Like
in condensed matter, such a nucleation is highly improbable. Indeed
a too small new vortex band is unstable. However when exceptionally, by chance, a
critical size is reached the new band becomes stable and will last
for an extremely long time. In combination with this growth, the three
jets move apart. It is striking to note that all nucleations ($2\rightarrow3$ transitions) have been observed at the edge of
westward jets, and all coalescences ($3\rightarrow2$ transitions) occurred at the edge of eastward
jets. 
This phenomenology is illustrated on an even clearer
way on Fig.~\ref{fig5} (a) that shows a typical zonal velocity evolution during
the nucleation of new jet and Fig.~\ref{fig5} (b) that shows jet coalescence.

The Arrhenius law, from thermodynamics and statistical physics, states
that transitions rates are proportional to $\lambda\propto\exp(-\Delta V/\alpha)$,
where $\Delta V$ is either a free energy, an entropy, or a potential
difference, and $\alpha$ is related to thermal or non-thermal noises.
This classical law describes transitions in many fields of physics,
chemistry, biology, statistical and quantum mechanics. Could it be
relevant to turbulence problems, extremely far from equilibrium? This
fascinating hypothesis has never been tested for turbulent flows because
this requires a huge number of rare transitions for different values
of $\alpha$, an impossible task without a rare event algorithm. The
validity of this hypothesis is suggested by the nucleation phenomenology.
Moreover, we have recently conjectured \cite{Bouchet_Nardini_Tangarife_2013_Kinetic_JStatPhys,bouchet2017fluctuations}
that the slow evolution of the zonally averaged part of the flow,
$U(y,t)=\int\mbox{d}x\,\mathbf{v}(x,y,t)$, may be described by an
effective equation
\begin{equation}
\frac{\partial U}{\partial\tau}=F(U)+\sqrt{\alpha}\sigma(U,\tau),\label{U}
\end{equation}
where $\tau=\alpha t$ is a rescaled time, $F(U)$, the average of
the divergence of the Reynolds stress (more precisely, \cite{Bouchet_Nardini_Tangarife_2013_Kinetic_JStatPhys} derived Eq. (\ref{U}) formally and proved that the hypothesis for the asymptotic expansion leading to $F(U)$ are self-consistent, while \cite{bouchet2017fluctuations} explained how to compute $\sigma(U,\tau))$). The classical Freidlin\textendash Wentzell
theory \cite{Freidlin_Wentzel_1984_book} describes large deviations
and rare transitions for Eq. (\ref{U}) for weak noises ($\alpha\ll1$).
From this theory, two main
consequences can be derived from Eq. (\ref{U}): first an Arrhenius law, and second a concentration of
transition paths close to a single path called instanton \cite{Bouchet_Nardini_Tangarife_Kazimierz,grafke2017long,grafke2015instanton}
(see \cite{Berhanu2007} for
an experimental observation in a magneto hydrodynamics turbulent flow,
and \cite{grafke2015instanton} for numerical results for Burger's
equation). In the following of this letter we will show that these two consequences are verified, giving further support to  Eq. (\ref{U}).

\begin{figure}
\centerline{\includegraphics[width=0.4\textwidth]{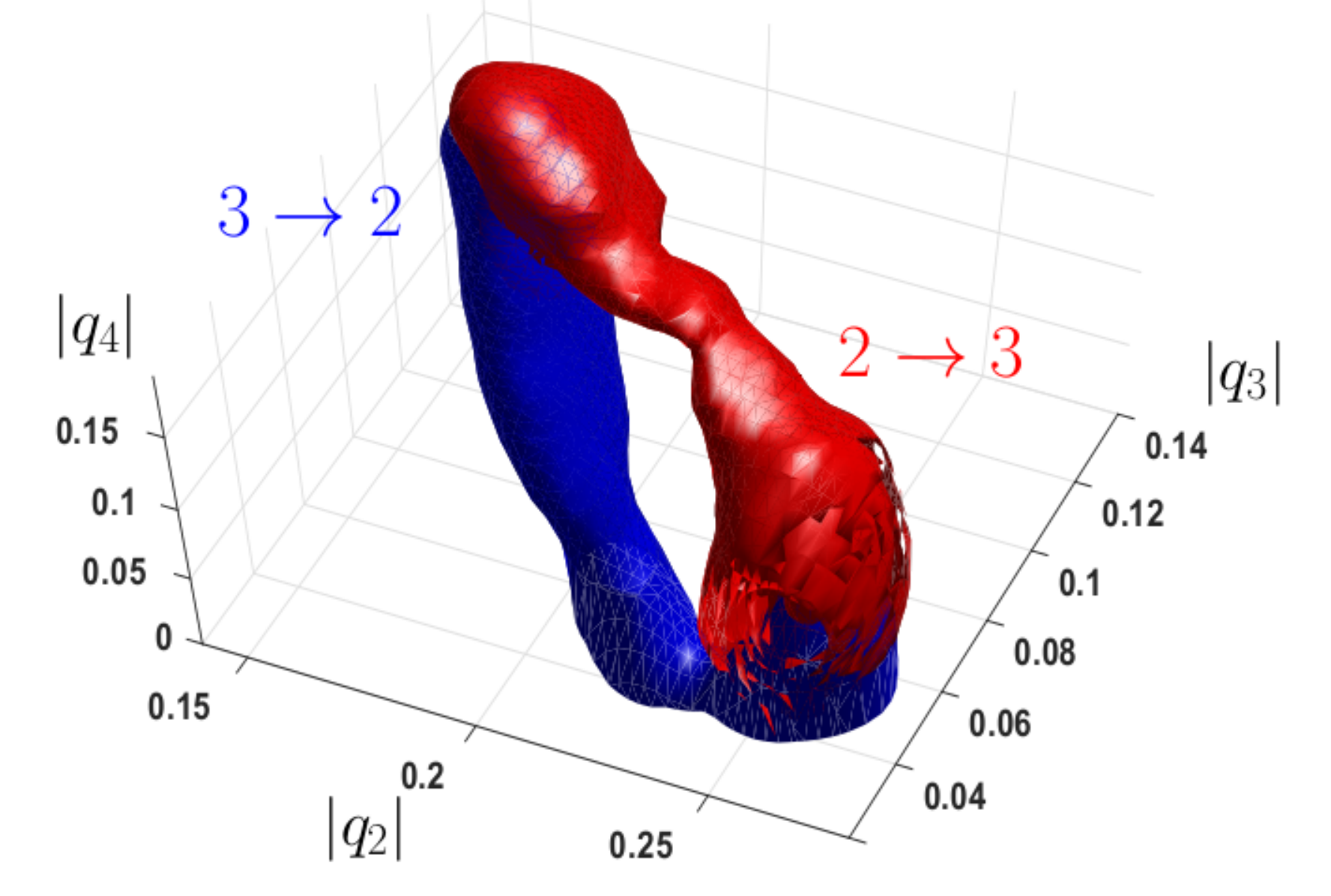}}
\caption{\textbf{Instantons:} The reactive tubes corresponding to the distribution
of transition paths for the $2\to3$ (red) and $3\to2$ (blue) transitions.
They illustrate the concentration of transition paths typical of an
instanton phenomenology (see the main text) ($\beta=5.26$ and $\alpha=1.2\cdot10^{-3}$).}
\label{fig6}
\end{figure}

\begin{figure}
\includegraphics[width=0.4\textwidth]{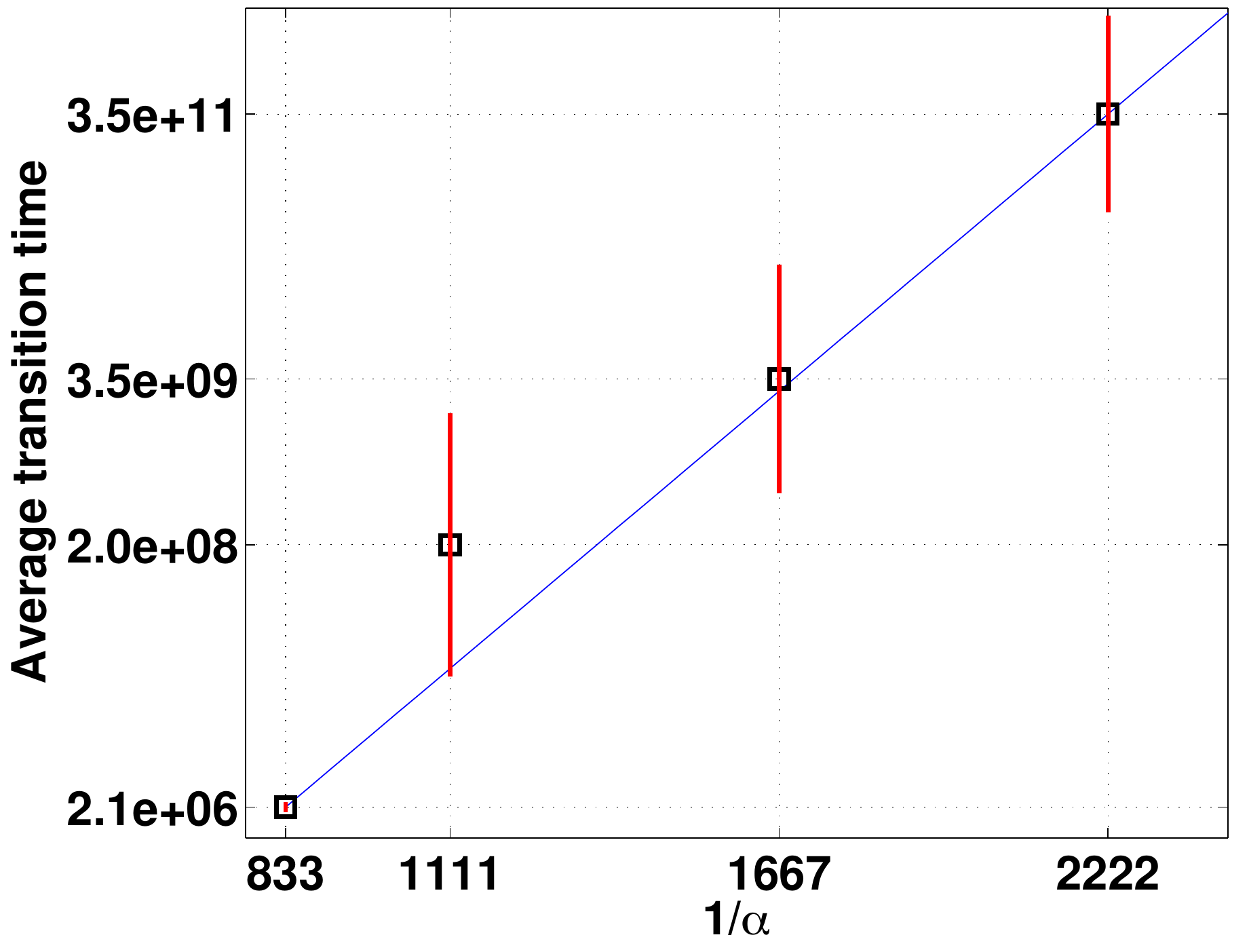}
\caption{\textbf{Arrhenius law. }Logarithm of the mean first transition
time $T$ from the two-jet to the three-jet attractors versus $1/\alpha$
($\beta=5.5$). This result suggests that mean transition times might
follow an Arrhenius law $T\propto\text{e}^{\Delta U/\alpha}$. (b)
}
\label{fig7}
\end{figure}

Using the adaptive multilevel splitting algorithm, we have been able
to collect thousands of transition paths. In Fig.~\ref{fig6}, $80\%$ of the\textbf{
}direct $2\to3$ transitions are inside the red tube, and $80\%$
of the direct\textbf{ }$3\to2$ trajectories are inside the blue tube,
in the reduced space\textbf{ }of observables $\left(\left|q_{2}\right|,\left|q_{3}\right|,\left|q_{4}\right|\right)$
(see Fig.~\ref{fig2}). This unambiguously illustrates the concentration of
transition paths close to an instanton. This is the first demonstration
of such a phenomenology from numerical simulations in a turbulent
flow. We stress the strong asymmetry between the $2\to3$ and $3\to2$
transition, which is expected for an irreversible dynamics of a turbulent
flow. We also study for the first time in a turbulent flow an Arrhenius
law, based on thousands of extremely rare transitions (see Tab.~\ref{tab1}).
Following the approach described in~\cite{CGLP11}
we compute the averaged transition time $\mathbb{E}(T)=1/\lambda$
for the $2\rightarrow3$ transitions (see Fig.~\ref{fig7}). Those data are
clearly compatible with an Arrhenius law $\log\mathbb{E}(T)\propto\Delta V/\alpha$. Viscosity effects are discussed in the Supplementary Material file.

\begin{table}
\begin{tabular}{|c|c|c|}
\hline $\alpha$ & AMS & DNS \\ \hline
$1.2\cdot 10^{-3}$&1.0d&15d \\ \hline
$0.9\cdot10^{-3}$&1.4d& 210d\\ \hline
$0.6\cdot10^{-3}$&2.2d& $\sim$51y\\ \hline
$0.45\cdot10^{-3}$&3.4d&  $\sim$2050y\\ \hline
\end{tabular}
\caption{CPU time (d: days, y: years) needed to obtain 1000 transition paths
using 200 processors for the adaptive multilevel splitting algorithm compared to direct numerical
simulation for different values of $\alpha$. }
\label{tab1}
\end{table}

The new use of the adaptive multilevel splitting algorithm for studying
rare transitions in a turbulent flow demonstrates for the first time
that thousands of vortices can self-organize and nucleate new structures
and trigger transitions. Like in condensed matter, the transition
paths concentrate close to instantons. Instantons may be used as precursors
stressing that the extremely rare transition became more probable. A very important future work will be the study of Jupiter's abrupt climate changes with models that are more realistic then the barotropic $\beta$-plane model \cite{Schneider_Liu_2009JAtS...66..579S}.
While it is unlikely that the same nucleation phenomenology should
hold for all possible turbulent transitions, the methodology developed
will prove useful to study many other transitions related to drastic
changes for climate, geophysical, astrophysical and engineering applications.
This opens a new range of studies impossible so far, and bring turbulent
phenomena in the realm of non-equilibrium statistical mechanics. Examples include the Kuroshio
current bistability, weather regime changes in meteorology, regime
transitions in the Sun superficial dynamics, astrophysical magnetic
field transitions, bistability in turbulent boundary layer detachments.

\bibliography{Fbouchet_2017_11,All-2018-09}

\section*{Acknowledgments}

The authors thank Corentin Herbert, Eric Vanden-Eijnden and Tobias
Grafke for useful discussions. The research leading to these results
has received funding from the European Research Council under the
European Union\textquoteright s seventh Framework Program (FP7/2007-2013
Grant Agreement No. 616811) (F. Bouchet). All three authors have equally contributed to the research and writing
of this letter.\\

This submission contains a Supplementary Material file and a Supplementary Video that features an extremely
rare nucleation of a new jet. Such a process has never been observed
before.\\

{\bf Please see next page for the Supplementary Materials.}


\newpage

\setcounter{page}{1}.

\onecolumngrid

\section{Supplementary Materials for the letter: A rare event algorithm links transitions in turbulent flows with activated nucleations, by F. Bouchet, J. Rolland and E. Simonnet}

\twocolumngrid

\paragraph{The beta-plane model for barotropic flows.}

All the results in this letter are based on the barotropic quasi-geostrophic
equations, with a beta plane approximation for the variation of the
Coriolis parameter. The equations in a doubly periodic domain $\mathcal{D}=[0,2\pi Ll_{x})\times[0,2\pi L)$
read
\begin{equation}
\partial_{t}\omega+\mathbf{v}\cdot\mathbf{\nabla}\omega+\beta_{d}v_{y}=-\lambda_f\omega-\nu_{n,d}\left(-\Delta\right)^{n}\omega+\sqrt{\sigma}\eta,\label{eq:barotropic-d}
\end{equation}
where $\mathbf{v}=\mathbf{e}_{z}\times\mathbf{\nabla}\psi$ is the
non-divergent velocity, $v_{y}$ the meridional velocity component,
$\omega=\Delta\psi$ and $\psi$ are the vorticity and the stream
function, respectively. $\lambda_f$ is a linear friction coefficient,
$\nu_{n,d}$ is a (hyper-)viscosity coefficient, and $\beta_{d}$
is the mean gradient of potential vorticity. $\eta$ is a white in
time Gaussian random noise, with spatial correlations
\[
\mathbf{E}\left[\eta({\bf r}_{1},t_{1})\eta({\bf r}_{2},t_{2})\right]=C({\bf r}_{1}-{\bf r}_{2})\delta(t_{1}-t_{2})
\]
that parameterizes the curl of the forces (physically due, for example,
to the effect of baroclinic instabilities or convection). The correlation
function $C$ is assumed to be normalized such that $\sigma$ represents
the average energy injection rate, so that the average energy injection
rate per unit of area (or equivalently per unit of mass taking into
account density and the layer thickness) is $\epsilon=\sigma/4\pi^{2}L^{2}l_{x}$.
These equations share the mathematical properties of the 2D Navier\textendash Stokes
equations and reduce to it when $\beta=0$.

The dynamics of large scale jet formation on Jupiter may be qualitatively
well understood within the framework of the barotropic quasi-geostrophic
equations with a $\beta$ plane approximation, although more refined
models are needed to understand their quantitative features \cite{Schneider_Liu_2009JAtS...66..579S}.
As the aim of this work is to make progress in the theoretical understanding
of turbulent flows, we consider the simple barotropic $\beta$ plane
model. Despite all its limitations, for instance the lack of dynamical
effects related to baroclinic instabilities, this model reproduces
the main qualitative features of the velocity profile and of the
jet spacing.\textbf{ }The aim of this study is to make a first study
of rare transitions for genuine turbulent flows which is directly
inspired by geophysical phenomena, rather than studying precisely
specific geophysical phenomena.

For atmospheric flows, viscosity is often negligible in the global
energy balance and this is the regime that we will study in the following.
Then the main energy dissipation mechanism is linear friction. The
evolution of the average energy (averaged over the noise realizations)
$E$ is given by
\[
\frac{dE}{dt}=-2\lambda_f E+\sigma.
\]
In a stationary state we have $E=E_{stat}=\sigma/2\lambda_f$, expressing
the balance between forces and dissipation. This relation gives the
typical velocity associated with the coherent structure $U\sim\sqrt{E_{stat}}/L\sim\sqrt{\epsilon/2\lambda_f}$.
We expect the non-zonal velocity
perturbation to follow an inviscid relaxation, on a typical time scale
related to the inverse of the shear rate $U/L$. Assuming that a typical
vorticity or shear is of order $s=U/L$ corresponding to a time $\tau=L/U$,
it is then natural to define a non-dimensional parameter $\alpha$
as the ratio of the shear time scale over the dissipative time scale
$1/\lambda_f$,
\[
\alpha=\lambda_f\tau=L\sqrt{\frac{2\lambda_f^{3}}{\epsilon}}.
\]
When $\beta$ is large enough, several zonal jets can develop in the
domain. An important scale is the so-called Rhines scale $L_{R}$
which gives the typical size of the meridional jet width:
\[
L_{R}=\left(U/\beta_{d}\right)^{1/2}=\left(\epsilon/\beta_{d}^{2}\lambda_f\right)^{1/4}.
\]

We write the non-dimensional barotropic equation using the box size
$L$ as a length unit and the inverse of a typical shear $\tau=L/U$
as a time unit. We thus obtain (with a slight abuse of notation, due
to the fact that we use the same symbols for the non-dimensional fields):
\begin{equation}
\partial_{t}\omega+\mathbf{v}\cdot\mathbf{\nabla}\omega+\beta v_{y}=-\alpha\omega-\nu_{n}\left(-\Delta\right)^{n}\omega+\sqrt{2\alpha}\eta,\label{eq:barotropic-1}
\end{equation}
where, in terms of the dimensional parameters, we have $\nu_{n}=\nu_{n,d}\tau/L^{2n}$,
$\beta=\beta_{d}L\tau$. Observe that the above equation is defined
on a domain $\mathcal{D}=[0,2\pi l_{x})\times[0,2\pi)$ and the averaged
stationary energy for $\nu_{n}\ll\alpha$ is of order one. Moreover the non dimensional number $\beta$ is equal to the square
of the ratio of the domain size divided by the Rhines scale. As a
consequence, according to empirical observations in numerical simulations, the number of jets approximately
scales like $\beta^{1/2}$ when $\beta$ is changed \cite{Dritschel_McIntyre_2008JAtS}.
We are interested in the strong jet regime, obtained for small values
of $\alpha$, which is relevant for Jupiter. All the computations
of this paper are performed with the parameters $\nu=1.5\ 10^{-8}$,
and using a stochastic force with a uniform spectrum in the wave number
band $|k|\in[14,15]$. We change the values of $\alpha$ and $\beta$
for different experiments, as explained in the letter.

\paragraph{The adaptive multilevel splitting algorithm and its validation.}

The sketch and caption of Fig.~\ref{fig3} explain the principle of the adaptive multilevel splitting algorithm. The cloning step for the discretization of a continuous time dynamics has to be more precisely defined. We denote $1$ the deleted trajectory, $1'$ the new trajectory, and $2$ the trajectory used for branching. $X_1(t)$ and $X_2(t)$ are the values of the model variables along the trajectories $1$ and $2$, respectively. The new trajectory $1'$ is equal to the trajectory $2$ up to the first time step $t_c$ such that $\mathcal{Q}(X_2(t_c))>\max_t \mathcal{Q}(X_1(t))$ \cite{Brehierlelievrerousset}. In the case of a discrete time
process, the branching level must therefore occurs at a discrete time strictly larger than $t_c$. Trajectory 1' is then computed as a numerical solution of (\ref{eq:barotropic-1}), with a new realization of the noise $\eta(t)$ for times larger than $t_c$  (we recall that $\eta$ is a random Gaussian field with two-point correlation function $C$; a new realization is a new sample of this random field). Any other branching choice adds a bias on the computed transition times. After $K$ iterations, the probability of the set of $N$ selected trajectories is $(1-1/N)^K$. If the user chooses to stop the algorithm at the iteration $K$ when $N$ trajectories reach the set $\mathcal{B}$, he obtains a set of $N$ reactive trajectories, which probability is $p = (1-1/N)^K$.
The averaged transition time is estimated as $\mathbb{E}(T)= \left(\frac{1}{p}-1\right) E_a + E_b$. The quantity $E_a$ is the averaged time to go from the boundary
$\partial {\cal A}$ of the set ${\cal A}$ to some hypersurface $\partial {\cal C}$ close to $\partial {\cal A}$ 
and going back to ${\cal A}$ without going to ${\cal B}$. This quantity is easily estimated by direct simulations since $\partial {\cal C}$ must be close to
$\partial {\cal A}$, The quantity $E_b$ is the averaged time to go from ${\cal A}$ to $\partial {\cal C}$ and then to ${\cal B}$ without first going back to ${\cal A}$.
The conditions under which this approximation is correct, and a justification, are discussed in details in \cite{CGLP11}.

A necessary condition for the adaptive multilevel algorithm to be
efficient is to have a good choice of the score function $\mathcal{Q}$
(see Fig.~\ref{fig3}). The optimal choice for $\mathcal{Q}$ is the probability
to reach the attractor $\mathcal{B}$ before reaching the attractor
$\mathcal{A}$ (the committor function). This committor function is
however unknown. The choice of $\mathcal{Q}$ should then be made
based on heuristic understanding of the transition dynamics. A bad
choice of $\mathcal{Q}$ can lead to the failure of the algorithm
to efficiently produce reactive trajectories, which is immediately
noticed by the user. A more subtle possible difficulty may occur when
several sets of transitions paths exists (bistability for the transition
paths, in the path space). Then several different score functions should
be used in order to compute independently different sets of transition
paths.

The reduced phase space is spanned by the moduli of the zonal Fourier
coefficients $q_{n}=\int\mbox{d}x\mbox{d}y\,\omega(x,y)\mbox{e}^{iny}/(2\pi)^{2}$
(see Fig~\ref{fig2}), with $n=2$, $n=3$ and $n=4$. We first approximate
their probability density functions (PDFs) by monitoring $\left|q_{2}\right|$,
$\left|q_{3}\right|$ and $\left|q_{4}\right|$ by direct numerical
simulation. The sets $\mathcal{A}$ and $\mathcal{B}$ (see Fig.~\ref{fig3})
correspond to some low-dimensional projections of the metastable states
having two and three eastern jets respectively, for some range of
values of $\left|q_{2}\right|$, $\left|q_{3}\right|$ and $\left|q_{4}\right|$
. This procedure defines two sets with disjoint compact support. The
support of $\mathcal{A}$ is inside the region $|q_{2}|\in[0.24,0.25],$
$|q_{3}|,|q_{4}|\in[0,0.05]$. The support of $\mathcal{B}$ is inside
a larger region $|q_{2}|\in[0.1,0.15]$, $|q_{3}|\in[0.20,0.30]$
and $|q_{4}|\in[0.1,0.2]$. We denote $\mathbf{M}$ a given point
in the reduced phase space with coordinates $|q_2|,|q_3|,|q_4|$,
and we define the reactive coordinate $\mathcal{Q}$ as $\mathcal{Q}(\mathbf{M})=d_{\mathcal{A}}/2d_{\mathcal{B}}$ if $d_{\mathcal{A}} < d_{\mathcal{B}}$, $\mathcal{Q}(\mathbf{M})=1-d_{\mathcal{B}}/2d_{\mathcal{A}}$ if $d_{\mathcal{A}} > d_{\mathcal{B}}$, and $\mathcal{Q}(\mathbf{M})=1/2$ if $d_{\mathcal{A}} = d_{\mathcal{B}}$;
where $d_{{C}}={\rm dist}(\mathbf{M},{C})$. The function $\mathcal{Q}$ is thus equal
to zero in the set $\mathcal{A}$ and one in the set $\mathcal{B}$.
This choice for $\mathcal{Q}$ is a very rough approximation of the committor isosurfaces but
gives rather good results with the adaptive multilevel splitting algorithm as demonstrated in this letter.

For $\alpha=1.2\cdot10^{-3}$, the only $\alpha$ value for which
transitions can be observed using direct numerical simulations, we
have verified that the transition paths obtained using the adaptive
multilevel splitting algorithm are qualitatively similar to the ones
observed in the direct numerical simulations. As seen on Fig.~\ref{fig2},
from the direct numerical simulation, only five $2\to3$ and five
$3\to2$ transitions have been observed. The time spent in the two
jet and three jet state is about $2.7\ 10^{6}$ and $2.1\ 10^{6}$
respectively over a total of $4.8\ 10^{6}$. This gives an extremely
rough estimate of the average transition time of order of $5.3\ 10^{5}$
for the $2\to3$ transition and $4.3\ 10^{5}$ for the $3\to2$ transitions,
respectively. As only 5 transitions have been observed, the uncertainty
of this estimate is huge, probably of about the same order of magnitude
as the estimate itself. Using the adaptive multilevel splitting algorithm
for the same parameters, with $N=1000$ clones for each realization
of the algorithm, and three realizations of the algorithm, we obtain
an estimate of the average transition time of $T\simeq5.8\cdot10^{5}$
for the $2\to3$. This is clearly compatible with what is observed
in the direct numerical simulation. We thus conclude that there is
a good qualitative agreement between the direct numerical simulations
and the adaptive multilevel splitting. A more quantitative agreement
can not be checked directly due to the prohibitive cost of direct
numerical simulations, but has been checked in models like the stochastic
Allen\textendash Cahn equation~\cite{Rolland_Bouchet_Simonnet_JStatPhys_2016},
for which explicit mathematical formula were available as a benchmark.

\paragraph{A remark about the instanton shape and the 3 jet attractor.}

On Fig.~\ref{fig6}, the red tube is a level set of the distribution of transition
paths in the reduced space $(|q_{2}|,|q_{3}|,|q_{4}|)$ for the $2\to3$
transition. It illustrates the concentration of transition paths typical
of an instanton phenomenology. The blue tube is the same for the $3\to2$
transitions ($\beta=5.26$ and $\alpha=1.2\cdot10^{-3}$). In the
Freidlin-Wentzell theory, for an irreversible dynamics like the Navier\textendash Stokes
equations and for a classical phenomenology with two attractors separated
by a saddle point, one expects the transition paths to form a figure
eight. This is not observed on Fig.~\ref{fig6}. One can also note on Fig.~\ref{fig6},
the large extension of the three jet attractor, related to the fact
that the observable $q_{3}$ has strong fluctuations for the three
jet state, as can be seen on Fig.~\ref{fig2}. Those fluctuations are associated
to the fact that for this value of $\beta$, the three jet state is
often asymmetric. For instance on Fig.~\ref{fig2}, for $550<\alpha t<1150$,
one clearly sees that two of the areas with negative values of the
vorticity (color blue), are smaller than the third. The extent of
this asymmetry changes in time, sometimes fast, sometimes very slowly,
suggesting a complex internal dynamics of the three jet attractor.
We believe that the absence of a figure height for the transition paths and
this apparently complex dynamics in the neighborhood of the three
jet attractor are deeply connected. A more precise explanation requires
a new study of the three jet attractor which will be the subject of
future works.

\paragraph{Hyperviscosity and the Arrhenius law. }

Fig.~\ref{fig7} features the logarithm of the transition time versus $1/\alpha$.
We note that for $1/\alpha=4\,500$, we have obtained a value of the
transition time lower than what should be expected according to the
Arrhenius law, by a factor of about two. We have also noticed that
for such small values of $\alpha$, hyperviscosity affects consequently
the energy balance, suggesting that hyperviscosity is too large for
reaching the zero hyperviscosity asymptotical regime. Lowering the
hyperviscosity would however require to use a better resolution for
the numerical simulation and represent a very subsequent numerical
effort that goes beyond the scope of this letter. Given these elements,
we conclude that while our results are compatible with the validity
of an Arrhenius law, a definitive numerical evidence of the validity
of an Arrhenius law would require better resolved numerical simulations.

\end{document}